\documentclass[letterpaper]{article}

\usepackage{natbib,alifeconf}  %% The order is important
\usepackage{url,hyperref,cleveref}
\usepackage{booktabs}

\title{Visualizing the Structure of Lenia Parameter Space}

\author{
    Barbora Hudcová$^{1*}$,
    František Dušek$^{2}$,
    Marco Tuccio$^{3}$, \and
    Clément Hongler$^1$ \\
    \mbox{}\\
    $^1$EPFL, Switzerland, *barbora.hudcova@epfl.ch \\
    $^2$CIIRC, CTU and FIT, CTU, Czech Republic \\
    $^{3}$ Universitat Pompeu Fabra, Spain
} 

\begin{document}

\maketitle

\begin{abstract}
    % Abstract length should not exceed 250 words
    Continuous cellular automata are rocketing in popularity, yet developing a theoretical understanding of their behaviour remains a challenge. In the case of Lenia, a few fundamental open problems include determining what exactly constitutes a soliton, what is the overall structure of the parameter space, and where do the solitons occur in it. In this abstract, we present a new method to automatically classify Lenia systems into four qualitatively different dynamical classes. This allows us to detect moving solitons, and to provide an interactive visualization of Lenia's parameter space structure on our website \url{https://lenia-explorer.vercel.app/}. The results shed new light on the above-mentioned questions and lead to several observations: the existence of new soliton families for parameters where they were not believed to exist, or the universality of the phase space structure across various kernels. 
\end{abstract}

\vspace{-0.5em}

% Choose one of: Full Paper, Summaries, or Late Breaking Abstracts 
Submission type: \textbf{Late Breaking Abstract}\\

\vspace{-0.5em}

% If sharing code / data, anonymize your repository and paste the link here.
% Example of anonymizing sevice for github: https://anonymous.4open.science/
% delete this line if not needed
\small
\noindent website: \url{https://lenia-explorer.vercel.app/}

\noindent data: \url{https://bit.ly/lenia-explorer-data}
 
\noindent code: \url{https://bit.ly/lenia-explorer-code}

\normalsize

\vspace{-0.5em}

\section{Introduction}
Lenia has recently become one of the most popular Alife models due to the beautiful patterns that frequently emerge in its dynamics \cite{chan2018lenia}. Despite the profound interest \cite{asymptotic_lenia, hamon2022learning, flow_lenia2, mace}, our theoretical understanding of Lenia remains very limited \cite{yevenko2025using, kojima2025glider}; we list a few open questions:
\begin{itemize}
\setlength\itemsep{.01em}
    \item What constitutes solitons, and how to automatically detect them?
    \item Where in the Lenia parameter space do solitons emerge?
    \item How does the emergence of solitons depend on the choice of initial configurations?
    \item What is the structure of the Lenia parameter space and how does the shape of the kernel affect it?
\end{itemize}

In this work we distinguish four qualitatively different types of Lenia's dynamical behaviour and we present a new method to automatically classify each system into one of them. This allows us to visualize the ``phase space'' for various Lenia kernels. The results can be interactively explored at our website, which helps to build an important understanding to the above mentioned questions, and which identifies soliton regions for a variety of kernels.

In \cite{yevenko2024classifying} the Lenia parameter space was shown to exhibit a fractal structure. The most thorough identification of solitons was done manually in \cite{chan2018lenia}, Figure 9 for one fixed kernel. 

\vspace{-0.3em}

\section{Method}
We study the classical variant of Lenia with a single channel; a kernel $K: \R \rightarrow [0,1]$, and a growth function $G: [0,1] \rightarrow [-1, 1]$. An $n \times n$ configuration at time $t$ is $A^t \in [0,1]^{n \times n}$; the Lenia update with a time-step $\Delta t \in \R_{\geq 0}$ at position $x$ is given by
$$A^{t+\Delta t}(x) = [A^t + \Delta t G(K * A^t(x))]_0^1$$ where $*$ denotes the convolution operation; for details see \cite{chan2018lenia}. In this work, we fix $\Delta t = 0.1$ and consider $G$ to be of the form $G_{\mu, \sigma} (x) = 2e^{-\frac{(x-\mu)^2}{2\sigma^2}}-1$; $\mu, \sigma  \in (0, 1)$. Thus, for a fixed kernel $K$, each Lenia system is characterized by the values $\mu$ and $\sigma$. Below, we define the four dynamical classes and our classification method.

\vspace{-0.2em}

\subsection{Dynamical Phases of Lenia Systems}
We fix a system with kernel $K$ and growth function parameters $\mu$ and $\sigma$. First, we define the dynamical phase of a given initial configuration. Then, we traverse the space of initial configurations to determine the system's overall behaviour.

\subsubsection{Classifying Initial Configurations}
For $A^0 \in [0,1]^{n \times n}$, we compute the trajectory $A^0, A^{\Delta t}, A^{2 \Delta t}, \ldots, A^{ T_{\mathrm{max}} \Delta t}$ with $T_{\mathrm{max}}  \approx 7000$ and assign to $A^0$ one of the following phases\footnotemark{}:\\
\includegraphics[scale=0.03]{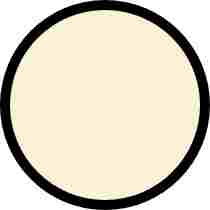} \textbf{Stable Phase: } The trajectory enters a loop.\\
\includegraphics[scale=0.03]{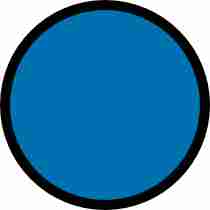} \textbf{Metastable Phase: } The trajectory does not enter a loop, but its center of mass stabilizes around its long-term mean.\\
\includegraphics[scale=0.03]{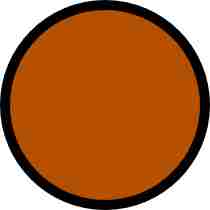} \textbf{Unclassified: } Neither of the two cases above.

\begin{figure}[htbp!]
    \centering
    \includegraphics[width=0.75\linewidth]{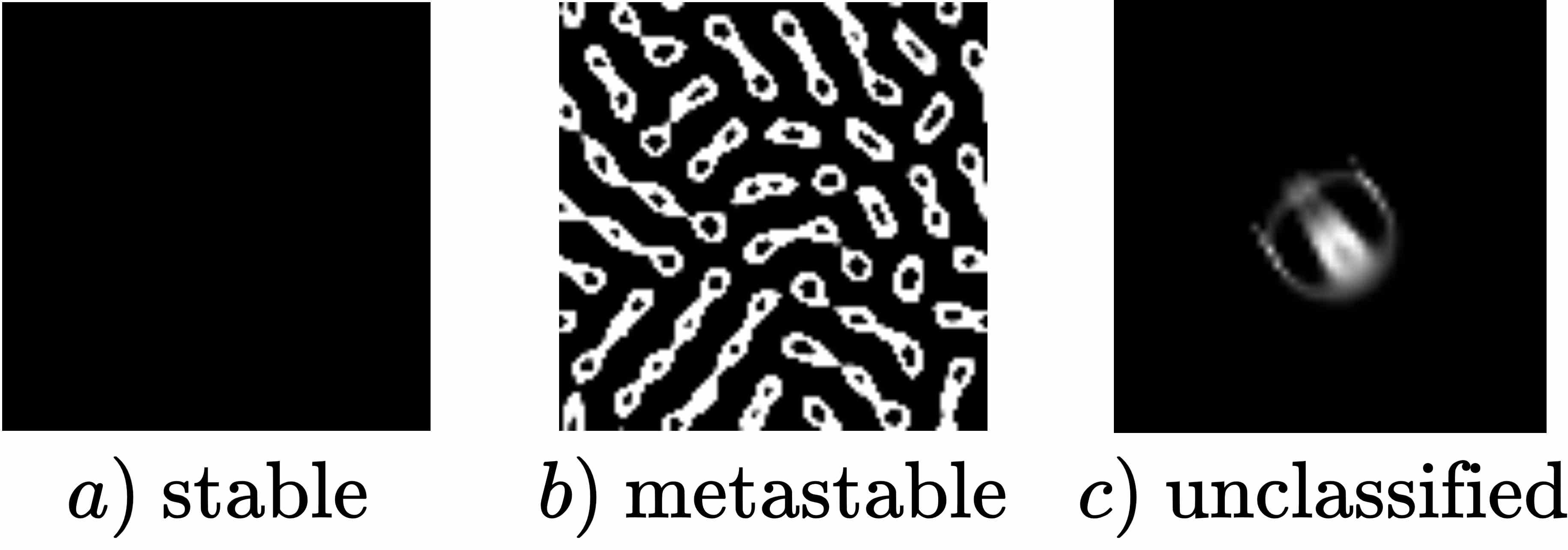}
    \vspace{-.7em}
    \caption{a) All activity dies out. b) Activity expands to the whole array. c) All moving solitons get ``unclassified''.}
    \label{fig:typical_phase_config}
    \vspace{-1em}
\end{figure}

A typical final configuration $A^{T_{\mathrm{max}}}$ of each phase is shown in Figure \ref{fig:typical_phase_config}.

Crucially, after analyzing over $10^5$ Lenia systems, we observed a strong correspondence between unclassified initial configurations and the emergence of moving solitons.

\subsubsection{Dynamical Classes of Lenia Systems}
\begin{figure}[htbp!]
    \centering
    \includegraphics[width=\linewidth]{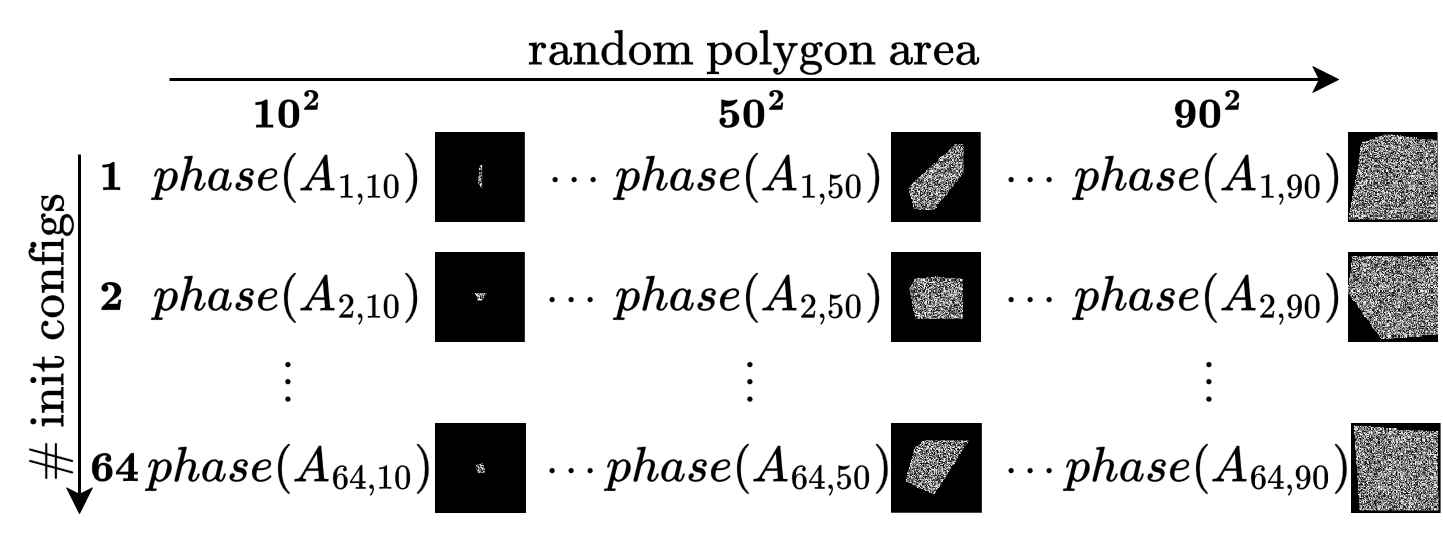}
    \vspace{-2em}
    \caption{Traversing the space of initial configurations by increasing the area of noise in the shape of random polygons.}
    \label{fig:traversing_configs}
\end{figure}

We now address a key question: how to choose initial configurations. The dynamics of many CAs depends strongly on this choice \cite{behrens2024dynamical}, and traversing Lenia’s initial-configuration space is essential to capture its full diversity. Following \cite{chan2018lenia}, we initialize with a noise patch (uniform random values in $[0,1]$) surrounded by zeros. To avoid bias toward specific shapes, we use the maximum-entropy principle: the shape of the patches is defined by random Voronoi polygons\footnotemark[\value{footnote}]. We traverse the configuration space by varying the size of noise patches; for a $100 \times 100$ grid, polygons of areas $10^2, 20^2, \ldots, 90^2$ are used. For each, we generate 64 configurations and determine their phases (Fig.~\ref{fig:traversing_configs}).

\footnotetext{Exact algorithmic details of the method and the choice of hyperparameters are provided in the documentation TODO.}

\vspace{-0.7em}

\vspace{-0.7em}

\begin{figure}[htbp!]
    \centering
    \includegraphics[width=0.95\linewidth]{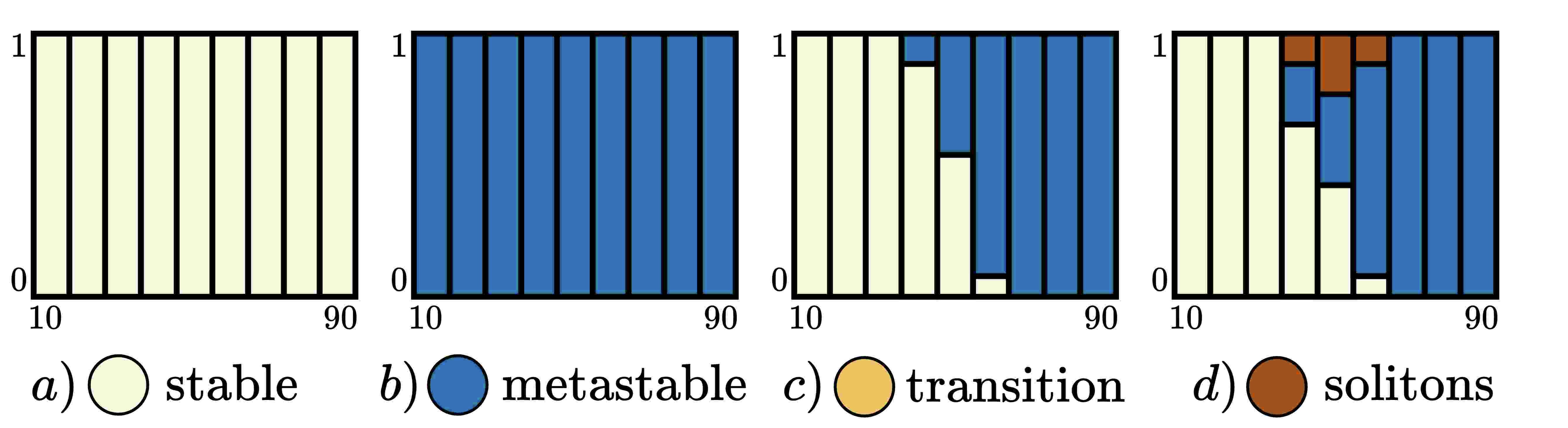}
    \vspace{-0.7em}
    \caption{x-axis: polygon sizes, y-axis: proportion of configurations in each phase. a) All configurations enter the stable phase. b) All configurations enter the metastable phase. c) A transition from stable to metastable phase as the patches of noise increase in size. d) A transition with solitons occuring around the transition region.}
    \label{fig:dynamical_phases}
    \vspace{-1em}
\end{figure}

We can summarize the matrix data of Fig. \ref{fig:traversing_configs} in a simple plot which, for each polygon area, shows the proportion of initial configurations classified into each phase. After analyzing a variety of systems, we observed four typical classes of dynamical behaviour summarized in Figure \ref{fig:dynamical_phases}.

To summarize the method, we described an algorithm that assigns to each Lenia system one of four classes of dynamical behaviour, allowing us to automatically detect systems (and initial configuration regions) where solitons occur.

\vspace{-0.5em}

\subsection{Results}

\begin{figure}[htbp!]
    \centering
    \includegraphics[width=0.95\linewidth]{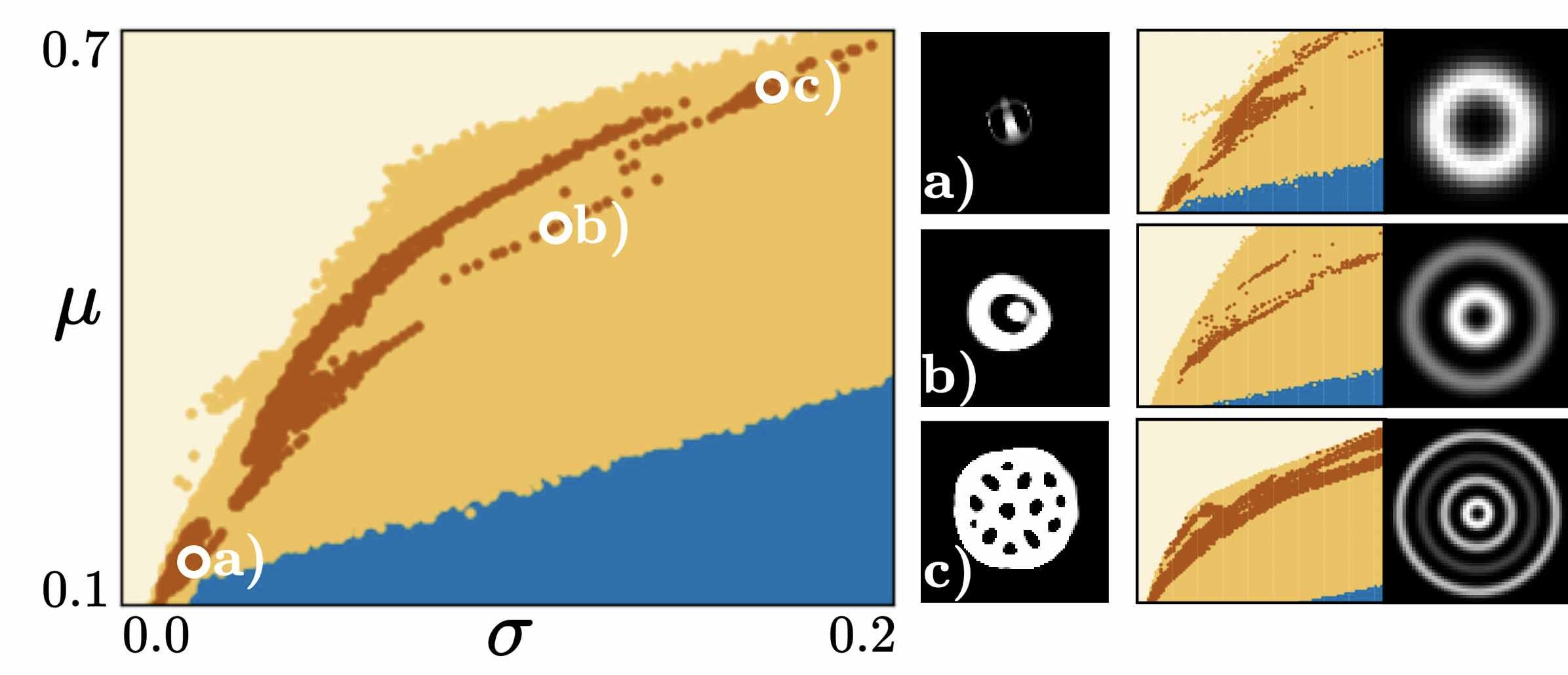}
    \vspace{-0.7em}
    \caption{(Left) ``Phase space'' of Lenia's dynamical classes for a fixed kernel while varying $\mu$ and $\sigma$. Dark orange region contains systems with emerging solitons, some of them showcased in the (middle). (Right) Analogous phase spaces for various kernel shapes depicted next to them with $0.1 \leq \mu \leq 0.5$ and $0.0 < \sigma \leq 0.1$.}
    \label{fig:final_result}
    \vspace{-1em}
\end{figure}

We have analyzed 8 kernels of various shapes. For each kernel, we vary the growth function parameters $\mu$ and $\sigma$, obtaining a $\mu-\sigma$ plane of around 10 000 systems, assigning a dynamical class to each. The complete results are made available on our interactive website\footnote{\url{https://lenia-explorer.vercel.app/}} and we encourage the reader to explore the fascinating range of emerging solitons. We illustrate the trend for the kernel $K(r) = e^{4 - \frac{1}{r(1-r)}}$ with radius $R=13$ in Fig. \ref{fig:final_result}.

In \cite{chan2018lenia}, Fig. 9, the same $\mu-\sigma$ plane was studied for an analogous kernel, though only for $\mu \leq 0.5$ $\sigma \leq 0.12$ as no solitons were believed to exist for larger values. Figure \ref{fig:final_result} thus shows the discovery of a new soliton families such as b) and c).

\vspace{-0.5em}

\subsection{Conclusion}
The dynamical classification of Lenia allowed us to make the following observations:
\vspace{-0.5em}
\begin{itemize}
    \item The method enabled automatic detection of solitons. \vspace{-0.7em}
    \item Solitons seem to occur in a very specific region of initial configurations, which typically lies in a transition region between the stable and metastable phase. \vspace{-0.5em}
    \item We discovered new soliton families, such as Figure \ref{fig:final_result} c). \vspace{-0.5em}
    \item We observe that a variety of kernels give rise to a qualitatively similar $\mu-\sigma$ ``phase space'', strinkingly resembling the phase transition of water. \vspace{-0.5em}
\end{itemize}

We believe that our results could help to give an analytical description between the stable and transition phase (more numerical results in \cite{papadopoulos2024looking}) as well as between the transition and metastable phase. Analytically characterizing the soliton region seems a more challenging, yet a very interesting goal. 

\section{Acknowledgements}
We would like to thank Bert Chan, Vassilis Papadopoulos, Ehsan Pajouheshgar, Eugène Bergeron, Franck Gabriel, and João Penedones for inspiring discussions. This work was funded by NCCR SwissMAP.

\begin{comment}
    We remark that Lenia systems in classes a) and b) exhibit minimal dynamical diversity, as the amount of randomness in the initial configuration does not seem to influence their overall behaviour. In contrast, systems in classes c) and d) showcase a certain sensitivity to the amount of noise. Remarkably, if we detect any solitons, they most typically occur around the phase transition from the stable to the metastable phase.
\end{comment}

\footnotesize
\bibliographystyle{apalike}
\bibliography{example} % replace by the name of your .bib file

@article{chan2018lenia,
  title={Lenia-biology of artificial life},
  author={Chan, Bert Wang-Chak},
  journal={arXiv preprint arXiv:1812.05433},
  year={2018}
}

@inproceedings{yevenko2024classifying,
  title={Classifying the fractal parameter space of the Lenia Orbium},
  author={Yevenko, Ivan},
  booktitle={Artificial Life Conference Proceedings 36},
  volume={2024},
  number={1},
  pages={14},
  year={2024},
  organization={MIT Press One Rogers Street, Cambridge, MA 02142-1209, USA journals-info~…}
}

@article{kojima2025glider,
  title={The Glider Equation for Asymptotic Lenia},
  author={Kojima, Hiroki and Yevenko, Ivan and Ikegami, Takashi},
  journal={arXiv preprint arXiv:2508.04167},
  year={2025}
}

@inproceedings{asymptotic_lenia,
  title={Introducing asymptotics to the state-updating rule in Lenia},
  author={Kawaguchi, Takako and Suzuki, Reiji and Arita, Takaya and Chan, Bert},
  booktitle={Artificial Life Conference Proceedings 33},
  volume={2021},
  number={1},
  pages={91},
  year={2021},
  organization={MIT Press One Rogers Street, Cambridge, MA 02142-1209, USA journals-info~…}
}

@article{hamon2022learning,
  title={Learning sensorimotor agency in cellular automata},
  author={Hamon, Gautier and Etcheverry, Mayalen and Chan, Bert Wang-Chak and Moulin-Frier, Cl{\'e}ment and Oudeyer, Pierre-Yves},
  year={2022}
}

@inproceedings{flow_lenia2,
  title={Flow-Lenia: Towards open-ended evolution in cellular automata through mass conservation and parameter localization},
  author={Plantec, Erwan and Hamon, Gautier and Etcheverry, Mayalen and Oudeyer, Pierre-Yves and Moulin-Frier, Cl{\'e}ment and Chan, Bert Wang-Chak},
  booktitle={Artificial Life Conference Proceedings 35},
  volume={2023},
  number={1},
  pages={131},
  year={2023},
  organization={MIT Press One Rogers Street, Cambridge, MA 02142-1209, USA journals-info~…}
}

@article{yevenko2025using,
  title={Using Dynamical Systems Theory to Quantify Complexity in Asymptotic Lenia},
  author={Yevenko, Ivan and Kojima, Hiroki and Nehaniv, Chrystopher L},
  journal={arXiv preprint arXiv:2508.02935},
  year={2025}
}

@misc{mace,
      title={MaCE: General Mass Conserving Dynamics for Cellular Automata}, 
      author={Vassilis Papadopoulos and Etienne Guichard},
      year={2025},
      eprint={2507.12306},
      archivePrefix={arXiv},
      primaryClass={nlin.CG},
      url={https://arxiv.org/abs/2507.12306}, 
}

@inproceedings{papadopoulos2024looking,
  title={Looking for Complexity at Phase Boundaries in Continuous Cellular Automata},
  author={Papadopoulos, Vassilis and Doat, Guilhem and Renard, Arthur and Hongler, Cl{\'e}ment},
  booktitle={Proceedings of the Genetic and Evolutionary Computation Conference Companion},
  pages={179--182},
  year={2024}
}

@article{behrens2024dynamical,
  title={Dynamical phase transitions in graph cellular automata},
  author={Behrens, Freya and Hudcov{\'a}, Barbora and Zdeborov{\'a}, Lenka},
  journal={Physical Review E},
  volume={109},
  number={4},
  pages={044312},
  year={2024},
  publisher={APS}
}

\end{document}